\documentclass{article}
\usepackage{spconf,amsmath,graphicx}
\usepackage{booktabs, multirow, amsfonts}
\usepackage{color, enumitem, url}

\title{Spatio-Temporal Attention in Multi-Granular Brain Chronnectomes for Detection of Autism Spectrum Disorder}
\name{James Orme-Rogers and Ajitesh Srivastava}
\address{ University of Southern California \\
\{ormeroge, ajiteshs\}@usc.edu}
\begin{document}
%
\ninept
\maketitle
\begin{abstract}
The traditional methods for detecting autism spectrum disorder (ASD) are expensive, subjective, and time-consuming, often taking years for a diagnosis, with many children growing well into adolescence and even adulthood before finally confirming the disorder. Recently, graph-based learning techniques have demonstrated impressive results on resting-state functional magnetic resonance imaging (rs-fMRI) data from the Autism Brain Imaging Data Exchange (ABIDE). We introduce IMAGIN, a mult\textbf{\underline{I}}-granular, \textbf{\underline{M}}ulti-\textbf{\underline{A}}tlas spatio-temporal attention \textbf{\underline{G}}raph \textbf{\underline{I}}somorphism \textbf{\underline{N}}etwork, which we use to learn graph representations of \textit{dynamic} functional brain connectivity (chronnectome), as opposed to \textit{static} connectivity (connectome). The experimental results demonstrate that IMAGIN achieves a 5-fold cross validation accuracy of 79.25\%, which surpasses the current state-of-the-art by 1.5\%. In addition, analysis of the spatial and temporal attention scores provide further validation for the neural basis of autism.
\end{abstract}
\section{Introduction}
\label{sec:intro}

Autism spectrum disorder (ASD) is considered to be a neurodevelopmental disorder resulting from abnormal development of brain connectivity. This atypical connectivity in autism has loosely been characterized by underconnectivity between distant brain regions and hyperconnectivity between neighboring ones \cite{holiga2019patients, anderson2011decreased}. ASD manifests as a wide variety of deficits in social communication and language acquisition, as well as restricted interests and repetitive behaviors \cite{monk2009abnormalities}.
Early detection is often cited as the most important factor in the successful treatment of autism. However, there are a number of significant barriers in the way of an accurate diagnosis. No quantitative metric or medical test currently exists that can quickly screen for autism, but instead patients must be subjected to intense observation of developmental behavior with medical specialists including neurologists, behavioral pediatricians, and psychiatrists.
We investigate a new paradigm in autism detection through the analysis of rs-fMRI data, which measures the hemodynamic response of the brain while the subject is at rest. By using blood flow as an indirect metric for brain activity, an intrinsic map of functional connectivity (FC) can be established for each subject.


The first work performed on this multi-site autism classification problem involved training generalized linear models and a leave-one-out classifier on FC matrices and phenotypic features from 7266 brain regions of interest (ROIs), achieving accuracy of 60.0\% \cite{nielsen2013multisite}. This remained as the state-of-the-art for years until a support vector classifier trained on a subset of ABIDE I (871 subjects who passed manual quality control inspections from three independent human reviewers) improved to 66.8\% accuracy on an atlas with only 84 ROIs \cite{abraham2017deriving}. As deep-learning methods gained popularity, new techniques such as MLPs, CNNs, autoencoders, and graph-based models brought a resurgent interest in the classification of ASD \cite{liu2021autism, horien2022functional}. 

Among graph-based models, the ASD detection problem can be formulated as both a node classification task and a graph classification task. In node classification, one graph is constructed for the entire population, with each node and its associated features corresponding to a single subject and their flattened FC matrices. Nodes are connected through weighted similarity metrics of phenotypic features, such as age, sex, handedness, and location of the scanning site. In graph classification, a graph is constructed for each individual in the ABIDE I dataset, with nodes representing ROIs and edges representing functional time series correlations between ROIs \cite{parisot2018disease}. Previously, the state-of-the-art in graph-based models was a population-based 16-layer Deep-GCN trained on FC matrices extracted from the 111-region Harvard-Oxford brain atlas. Using the same ABIDE I subset of 871 subjects as described in \cite{abraham2017deriving}, they achieved a 10-fold cross-validation accuracy of 73.7\% \cite{cao2021using}. This was recently surpassed, however, by an ensemble graph-based learning method that classified each individual with 75.9\% accuracy according to a majority vote from six different brain atlases, ranging from 110 to 200 ROIs \cite{wang2022mage}. Both of these graph-based models only used static FC information, where pairwise correlations between regions are calculated using the entire duration of the fMRI scan. This neglects the fact that our brains are dynamic systems, with functional connectivity fluctuating over short periods of time, even when at rest. This has led to the distinction between the “connectome” and the “chronnectome” - with the analysis of time-varying connectivity networks being established as the next frontier in fMRI data discovery \cite{calhoun2014chronnectome, preti2017dynamic}. The current SOTA in dynamic graph-based models only achieves 70.7\% 10-fold cross-validation accuracy, with a notable limitation being that it measures temporal functional connectivity with a single frame at a time, rather than with a sliding-window approach \cite{wang2021graph}.

Our key contributions to this autism detection problem are as follows: (i) A novel multi-granular modeling technique to capture fine-grained local connectivity (voxels) as well as coarse-grained connectivity (atlas regions) across the dynamic connectome, (ii) A multi-atlas individual-level graph-based ensemble learning technique that incorporates \textit{dynamic} functional connectivity (dFC) data with phenotypic features, and (iii) We integrate spatial and temporal attention into a Graph Isomorphism Network (GIN) to provide a measure of explainability for our methods - not only to indicate the most discriminative brain regions for this detection problem, but to also show how variations in connectivity can be quantified over time. In evaluating the patterns of FC between subjects with ASD and healthy controls (HC), we achieve state-of-the-art ASD classification performance (79.25\% accuracy) over all previous methods, successfully combining a graph-based multi-atlas model architecture with dynamic functional connectivity and phenotypic features.

\section{Materials and Methodology}

\subsection{The ABIDE I Dataset}

The ABIDE (Autism Brain Imaging Data Exchange) dataset provides resting-state fMRI data of subjects with autism and healthy controls collected from across 17 international sites. Previous research has shown that when classifying scans from a single site, accuracy is quite high, but drops drastically when scans across multiple sites are considered, perhaps due to scanning parameter differences between sites \cite{nielsen2013multisite}. In the tradition of previous ABIDE research referenced in this paper, we only use data from the ABIDE I Preprocessed Connectomes Project \cite{craddock2013neuro} preprocessed through the Configurable Pipeline for the Analysis of Connectomes (CPAC), with band-pass filtering and no global signal regression. Following \cite{abraham2017deriving}, we take the 871 subjects that passed manual quality control checks from three expert human reviewers and further prune the data with less than 150 time steps to a final total of 809 samples. The resulting class distribution is 383 subjects with ASD to 426 healthy controls.

\subsection{Dynamic Graph Construction}


Traditionally, rs-fMRI analysis was done under the outdated assumption that the activity of the brain remains stagnant throughout the duration of the scan and  the functional connectivity between brain regions is static. Research has shown, however, that even throughout the relatively short duration of an MRI scan, the brain is a \textit{dynamic} entity, displaying measurable changes in functional connectivity over time periods of milliseconds, seconds, and minutes \cite{vidaurre2017brain}.

To construct a sequence of dFC graphs, we utilize the popular sliding window approach \cite{calhoun2016time}. For a given window length $\Gamma$, voxel timeseries are extracted, averaged, and standardized from each brain atlas ROI to form an aggregate ROI timeseries $\textbf{\textit{r}}_{i}(t) \in \mathbb{R}^{\Gamma}$. Pairwise correlation coefficients between ROI timeseries are calculated to construct FC matrices and the top $\alpha$-percentile correlations are thresholded to produce a sparse, weighted adjacency matrix $\textbf{\textit{A}}(t)$.



The window is then continually shifted by stride $S$ until the end of the fMRI scan to construct $T = \lfloor T\textsubscript{max} - \Gamma/S \rfloor$ total windows, resulting in a sequence of sparse dFC matrices $\textbf{\textit{A}}_\textsubscript{dyn} = (\textbf{\textit{A}}(1),...,\textbf{\textit{A}}(T))$.

The node features $\textbf{x}_n(t)$ corresponding to each ROI $n$ in window $t$ are a concatenation of three sets of features: (i) one-hot encoded ROI vectors, (ii) encoded timestamps from a Gated Recurrent Unit (GRU) that receives as input the ROI timeseries from the beginning of the fMRI scan until the end of the sliding window following \cite{kim2021learning}, and (iii) average clustering coefficient and density of intra-regional voxel connectivity, which is further detailed in section \ref{subsec:graph-structural}.
Thus, for each sliding window $t$, the dynamic input graphs $G(t) = (V(t), E(t))$ are defined by the temporal node features $\textbf{\textit{X}}(t)$ and adjacency matrices $\textbf{\textit{A}}(t)$, forming the vertex set $V(t)$ and edge set $E(t)$, respectively.

\begin{table*}[t]
\centering
\begin{tabular}{@{}r|cccc|ccc@{}}
     \multirow{2}{*}{Model} & \multicolumn{4}{c|}{ABIDE I} & \multirow{2}{*}{Type of FC} & \multirow{2}{*}{Phenotypic} & \multirow{2}{*}{Multi-Atlas} \\
      & Accuracy (\%) & Precision (\%) & Recall (\%) & AUROC & & & \\ \hline
     
     ST-GCN & 68.4 & 64.4 & 69.8 & 0.705 & Dynamic & No & No \\
     cGCN & 70.7 & - & - & - & Dynamic & No & No \\
     Deep-GCN & 73.71 & 74.58 & 66.51 & 0.7520 & Static & Age, Site & No \\
     MAGE & 75.86 & 71.53 (sp) & 79.24 & 0.8314 & Static & Sex, Site & 6 \\
     \hline
     SVM+RFE & 76.63 & 74.27 (sp) & 78.63 & 0.8274 & Dynamic & Age, Sex, Site & No \\
     SVM+MTFS & 76.7$\pm$2.7 & 79.9$\pm$3.8 (sp) & 72.5$\pm$3.2 & 0.81$\pm$.031 & Dynamic & No & No \\
     \hline
     MISO-DNN & 77.73$\pm$4.26 & 76.73$\pm$4.11 & 77.16$\pm$3.72 & - & Static & No & 3 \\ \hline
     $e$-STAGIN\textsubscript{AAL} & 74.29$\pm$3.60 & 73.09$\pm$1.61 & 78.26$\pm$3.65 & 0.7996$\pm$.019 & Dynamic & No & No \\
     $e$-STAGIN\textsubscript{CC200} & 74.03$\pm$2.21 & 74.58$\pm$2.69 & 73.77$\pm$1.73 & 0.7840$\pm$.016 & Dynamic & No & No \\
     $e$-STAGIN\textsubscript{Sch} & 75.81$\pm$1.70 & 78.03$\pm$2.34 & 75.72$\pm$1.94 & 0.8112$\pm$.029 & Dynamic & No & No \\
     \hline
     MAGIN & 78.12$\pm$1.91 & 78.37$\pm$2.11  & \textbf{79.55}$\pm$\textbf{1.62} & 0.8572$\pm$.020 & Dynamic & Age, Sex, Site & 3 \\
     IMAGIN & \textbf{79.25}$\pm$\textbf{2.33} & \textbf{81.03}$\pm$\textbf{3.47} & 79.06$\pm$0.89 & \textbf{0.8644}$\pm$\textbf{.024} & Dynamic & Age, Sex, Site & 3 \\
\end{tabular}
\caption{Comparison of Cross Validation Results on ABIDE I Dataset. Specificity scores (sp) are used in place of precision when it was not reported in previous research.}
\label{results}
\end{table*}

\begin{figure}[t]
    \centering
    \includegraphics[width=\linewidth]{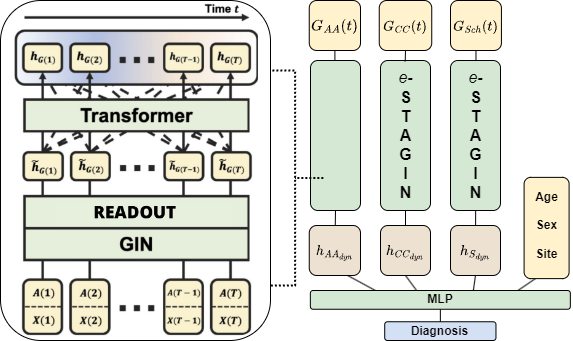}
    \caption{IMAGIN ensemble learning architecture.}
    \label{IMAGIN}
\end{figure}

\section{IMAGIN}
\label{sec:imagin}

\subsection{Multi-Granular Graph-Structural Information}
\label{subsec:graph-structural}

ASD manifests as an abnormal pattern of connectivity both throughout the brain at-large and within localized regions of interest \cite{rane2015connectivity}. However, no previous methods have been able to effectively capture network dynamics at the level of individual voxels. 



We propose a novel method to extract local intra-region connectivity information at the scale of individual voxels, in addition to the standard technique of computing pairwise correlations between brain parcellations across greater spatial distances.

\begin{enumerate}
    \item Extract standardized timeseries of individual voxels 
    within each region $n$ of the specified brain atlas.
    \item Generate a $k$-nearest neighbors graph, where $k$ is set to $\sqrt{\text{\# of voxels in } n}$, that reduces the number of connections to the top-$k$ signed correlation coefficients for each ROI $n$. This can be achieved with the following distance function derived from the standard Euclidean distance:
    \begin{equation}
        d(\textbf{x}^{\prime}, \textbf{y}^{\prime}) = 2 - \Bigl|(\sum_{t \in T} \delta(\textbf{x}_{t}^{\prime}, \textbf{y}_{t}^{\prime}))^2 - 2\Bigr|,
    \end{equation}
    where
        $\delta(\textbf{x}_{t}^{\prime}, \textbf{y}_{t}^{\prime}) = |\textbf{x}_{t}^{\prime} - \textbf{y}_{t}^{\prime}|^2$.
    \item Set the edge weights between voxels to their pairwise Pearson correlation coefficients:
    \begin{equation}
        r_{{\textbf{x}^{\prime}\textbf{y}^{\prime}}} = \frac{\mathrm{cov}(\textbf{x}^{\prime},\textbf{y}^{\prime})}{\sigma_{\textbf{x}^{\prime}}\sigma_{\textbf{y}^{\prime}}} \in [-1, 1],
    \end{equation}
    where $\mathrm{cov}$ denotes the covariance.
    \item Calculate the graph-structural features of density and average clustering coefficient for the weighted $k$-nearest neighbors graph. The density for undirected graphs is defined by
        $d = \frac{2m}{n(n-1)}$,
    where $n$ is the number of nodes and $m$ is the number of edges in the $k$-NN graph $G$. The average clustering coefficient for signed weighted graphs \cite{costantini2014generalization} is defined as
        $C = \frac{1}{n}\sum_{u \in G} c_u$,
    where $n$ is the number of nodes in $G$. The clustering $c_u$ is defined as the geometric average of the subgraph edge weights,
    \begin{equation}
        c_u = \frac{1}{k_u(k_u-1))}\sum_{vw} (\hat{w}_{uv} \hat{w}_{uw} \hat{w}_{vw})^{1/3},
    \end{equation}
    where $k_u$ is the degree of $u$ and the edge weights $\hat{w}_{uv}$ are normalized by the maximum weight $\Delta(G)$ of the network: $\hat{w}_{uv} = w_{uv}/\Delta(G)$.
\end{enumerate}

\subsection{Model Architecture}

At the core of our model lies a modified Graph Isomorphism Network (GIN) architecture from \cite{kim2021learning} that is augmented with spatial and temporal attention. Previously, this Spatio-Temporal Attention Graph Isomorphism Network (STAGIN) model was created to solve the problem of learning dynamic brain connectome graph representations while also providing temporal explainability. This can be formalized as mapping a sequence of graphs $G_{\textsubscript{dyn}} = (G(1),...,G(T))$ with $T$ timesteps to an embedding $\textbf{\textit{h}}_{G_\textsubscript{dyn}}$. This embedding is generated in a two-step process through the combination of a GIN and Transformer encoder with self-attention.

Graph Isomorphism Networks \cite{xu2018how} are a variant of traditional GNNs designed specifically for the graph classification task. Omitting timepoint notation (t), we modify the GIN layer to support edge weights $e_{ij}$:
\begin{equation}
    \textbf{\textit{h}}_{i}^{(k)} = \texttt{MLP}^{(k)}\Bigl( \textbf{\textit{x}}_{i}^{(k-1)} + \sum_{j \in \mathcal{N}(i)} e_{ij}\textbf{\textit{x}}_{j}^{(k-1)}\Bigr),
\end{equation}
where $\textbf{\textit{h}}_{i}^{(k)}$ is the representation of the node $\textbf{\textit{x}}_{i}$ at layer $k$. We refer to this modified architecture as $e$-STAGIN. 

The main contribution to the standard GIN architecture that we use in $e$-STAGIN is a novel attention-based adaptive READOUT mechanism that pools a spatially-attended global graph-level representation $\tilde{\textbf{\textit{h}}}_{G}$ from each of the node embeddings $\textbf{\textit{h}}_{i}^{(k)}$.

Before applying the attention mechanism, the average of the node embeddings in each layer $k$ defined by
\begin{equation}
    \textbf{\textit{h}}_{G(t)}^{(k)} = \frac{1}{N} \sum_{i \in G(t)} \textbf{\textit{h}}_{i}^{(k)},
\end{equation}
are concatenated into the global graph embedding $\textbf{\textit{h}}_{G(t)}$.

According to the procedure in \cite{kim2021learning}, the attention function $s(\textbf{\textit{h}}_{G(t)})$ produces the node coefficient vectors $\textbf{\textit{z}}_\textsubscript{space}^{(k)} \in [0,1]^{N}$ indicating the relative level of importance of each node in layer $k$. The spatially-attended global graph representation $\tilde{\textbf{\textit{h}}}_{G}^{(k)}$ is acquired by scaling each node embedding $\textbf{\textit{h}}_{i}^{(k)}$ by its corresponding $\textbf{\textit{z}}_\textsubscript{space}^{(k)}$ coefficient.


For attention across time, the spatially-attended graph embeddings $\tilde{\textbf{\textit{h}}}_{G_\textsubscript{dyn}} = (\tilde{\textbf{\textit{h}}}_{G(1)},...,\tilde{\textbf{\textit{h}}}_{G(T)})$ are fed as inputs to a single-headed Transformer encoder, which in turn computes the dynamic graph embeddings $\textbf{\textit{h}}_{G_\textsubscript{dyn}}^{(k)}$ from each layer as a summation over time. The self-attention weights $\textbf{\textit{Z}}_\textsubscript{time} \in [0,1]^{T \times T}$ can now be interpreted as a measure of the temporal importance of each dynamic graph throughout the length of the scan \cite{hao2021self}. The final output of the $e$-STAGIN model is the dynamic graph embedding, 
\begin{equation}
    \textbf{\textit{h}}_{G_\textsubscript{dyn}} = [\textbf{\textit{h}}_{G_\textsubscript{dyn}}^{(1)} || ... || \textbf{\textit{h}}_{G_\textsubscript{dyn}}^{(K)}],
\end{equation}
a concatenation of each spatially and temporally-attended dynamic graph embedding across all $K$ $e$-STAGIN layers.

We further expand upon the $e$-STAGIN architecture to create IMAGIN (mult\textbf{\underline{I}}-granular, \textbf{\underline{M}}ulti-\textbf{\underline{A}}tlas spatio-temporal attention \textbf{\underline{G}}raph \textbf{\underline{I}}somorphism \textbf{\underline{N}}etwork), a multimodal ensemble learning architecture that integrates the outputs of three separate multi-granular $e$-STAGIN modules with phenotypic data to produce a diagnosis.

\begin{figure*}[t]
\centering
\includegraphics[width=\textwidth]{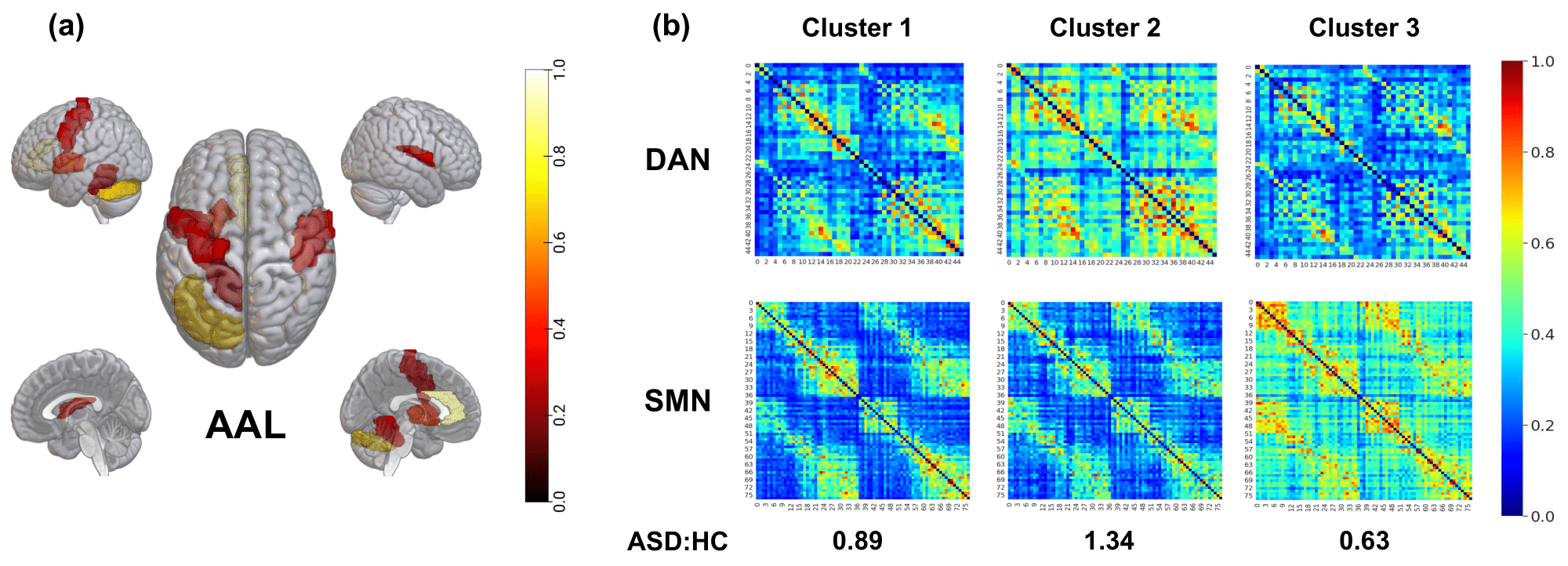}
\caption{(a) Top-5 percentile spatial attention regions for AAL atlas in IMAGIN. (b) $k$-means clustering is used to measure temporally-attended connectivity variations between the Dorsal Attention (DAN) and Somatomotor Networks (SN) in ASD and HC. 
}
\label{SattTatt}
\end{figure*}


Each $e$-STAGIN module output from individual brain atlases - $\textbf{\textit{h}}_{AAL_\textsubscript{dyn}}$, $\textbf{\textit{h}}_{CC200_\textsubscript{dyn}}$, and $\textbf{\textit{h}}_{Sch_\textsubscript{dyn}}$ - is fed as input along with phenotypic features (age, sex, scanning site) to a linear layer and the final ASD diagnosis is predicted with a $\mathrm{softmax}$ classifier. The full IMAGIN architecture is illustrated in Figure \ref{IMAGIN}.

\section{Results}

We utilize a stratified 5-fold cross-validated training procedure with a data split of 65\% training, 15\% validation, and 20\% testing during hyperparameter tuning. The stratification occurs over both class (ASD and HC) and site (one of 17 scanning locations), in attempt to minimize the effects of the different scanning parameters on our dataset distribution.

Due to the combinatorial explosion of hyperparameter tuning in the multi-atlas learning regime, optimal hyperparameters used for each $e$-STAGIN module within IMAGIN were found by training a single $e$-STAGIN module with individual atlases. 
With the optimal hyperparameters found, we train IMAGIN for 50 epochs on the same data splits used for each $e$-STAGIN module to compare performance under reproducible conditions. In addition, to evaluate the effectiveness of our multi-granularity feature engineering method, we train IMAGIN without such node features (MAGIN).


Since each $e$-STAGIN module accepts only a fixed-length sequence of graphs, we stochastically slice each subject's fMRI timeseries into a continuous sequence of 120 timesteps (4 minutes), with the minimum length scan in our subsample of the ABIDE-I dataset being 150 timesteps (5 minutes). We leave the tuning of this dynamic sequence length to future work. The validation and test sets are subsampled equivalently when evaluating model performance. In order to ensure reproducibility, this random subsampling is set with the same seed used to stratify our cross-validation folds.


We train using cross entropy loss with the Adam optimizer. All experiments were performed on a system composed of an Intel Core i9-9900K processor, 64 GB of RAM, and an NVIDIA GeForce RTX 3090 GPU.
Our code is publicly available\footnote{\url{https://github.com/jamiesonor/imagin}} for reproducibility.
The 5-fold cross-validation results of our methods are presented in Table \ref{results}, with classification performance compared to previous results. We achieve state-of-the-art performance across all model categories, beating the top graph-based models: ST-GCN \cite{cao2021temporal}, cGCN \cite{wang2021graph}, Deep-GCN \cite{cao2021using}, and MAGE \cite{wang2022mage}; dFC models: SVM+RFE \cite{karampasi2021informative} and SVM+MTFS \cite{liu2020improved}; and multi-atlas models: MISO-DNN \cite{epalle2021multi}. In addition, the introduction of multi-granular graph-structural node features results in a clear increase in accuracy between MAGIN and IMAGIN (+1.13\%).


\subsection{Spatio-Temporal Attention Analysis}

The spatial attention $\textbf{\textit{z}}_\textsubscript{space}^{(k)}(t)$ of each region is averaged across each layer and throughout time to create an aggregated measure of the importance of each node $\textbf{\textit{z}}_\textsubscript{space} \in [0,1]^{N}$. The top-5 percentile ROIs for the AAL atlas are shown in Figure \ref{SattTatt}(a). In particular, the putamen of the striatum complex is selected, as well as cerebellar regions in the left hemisphere, which have previously been established as loci for the neural basis of repetitive behaviors in ASD \cite{d2015cerebro, zhang2022connectivity}. Furthermore, this analysis was reproduced for the spatial attention results of the Schaefer atlas separated into seven distinct Yeo intrinsic connectivity networks (ICNs) \cite{yeo2011organization}. Both the Salience Network (SN) and the Dorsal Attention Network (DAN), responsible for the collaborative ventral and dorsal streams of attention, are especially attended to. The SN is the basis for the ``bottom-up" view of attention, which draws attention to unexpected or anomalous stimuli, while the DAN, or ``top-down" view, is the locus of directed, conscious allocation of focus. Attention deficits are hallmark symptoms of autism, with atypical development of the SN and DAN being cited as reasons for the impairment \cite{farrant2016atypical}. Finally, a large number of regions from the Default Mode Network (DMN) and from sensory regions of the Somatomotor Network (SMN) are highlighted, which likely correspond to the social and sensory processing deficits associated with ASD, respectively \cite{monk2009abnormalities, marco2011sensory}.


In analyzing the temporal attention results, we use $k$-means clustering to group the dFC information of all subjects. We effectively partition the set of fMRI timeseries into $k$ ``brain activity states." Following the methods of \cite{tagliazucchi2012criticality}, we take the set of all attended timepoints $\Bar{T}$ such that each attended timepoint is at least one standard deviation greater than the average temporal attention score $\textbf{\textit{z}}\textsubscript{time} \in [0,1]^{T}$ across each layer, and we set the number of clusters $k$ to 3. Removing the unattended timepoints from each subject's sequence of dFC $\textbf{\textit{A}}(t)$, we assign each subject to their closest cluster. Using the temporal attention scores from the Schaefer atlas, we examine the temporal states of the Dorsal Attention Network (DAN) and Somatomotor Network (SMN) in particular. We calculate the ratio of ASD subjects to healthy controls (HC) assigned to each cluster. It can be observed in Figure \ref{SattTatt}(b) that for the clusters in which there is hyperactivity in the DAN compared to the SMN, the ratio of ASD to HC is significantly greater than one (1.34). Conversely, when there is hypoactivity in the DAN compared to the SMN, the ratio of ASD to HC is less than one (0.63). This validates previous findings that discovered significant underconnectivity in sensorimotor regions and heightened connectivity in prefrontal and parietal regions among subjects with ASD \cite{holiga2019patients, anderson2011decreased}.

\section{Conclusion}

Overall, we introduce the multi-granular multi-atlas ensemble learning method IMAGIN, and demonstrate state-of-the-art results on this ASD detection task. For future work, we look to exploit the fact that FC patterns change between children and adults \cite{dosenbach2010prediction} in order to weight correct diagnoses for children more heavily. The age distribution of the ABIDE-I dataset is likely much older than if this method were to be tested in a clinical setting, so new techniques would need to be investigated to solve this domain shift problem. With that said, we hope to integrate this automated diagnostic system as part of a new paradigm in autism detection, and bring about a critical reduction in the time to diagnosis.

\bibliographystyle{IEEEbib}
\bibliography{biblio}

\end{document}